\documentstyle[12pt]{article}
\newcommand{\be}{\begin{equation}}
\newcommand{\bea}{\begin{eqnarray}}
\newcommand{\eea}{\end{eqnarray}}
\newcommand{\ba}{\begin{array}}
\newcommand{\ea}{\end{array}}
\newcommand{\ee}{\end{equation}}

\expandafter\ifx\csname mathbbm\endcsname\relax

\else

\fi
\begin{document}
\begin{titlepage}
\hfill
\vbox{
    \halign{#\hfil         \cr
           hep-th/0203018 \cr
           IPM/P-2002/004 \cr
           SU-ITP-02/10 \cr
           } 
      }  
\vspace*{20mm}
\begin{center}
{\Large {\bf The PP-Wave Limits of Orbifolded $AdS_5\times S^5$ }\\ }

\vspace*{15mm}
\vspace*{1mm}
{Mohsen Alishahiha$^a$ and Mohammad M. Sheikh-Jabbari$^b$}

\vspace*{1cm}

{\it $^a$ Institute for Studies in Theoretical Physics and Mathematics (IPM)\\
P.O. Box 19395-5531, Tehran, Iran\\ 
\vspace*{1mm}
$^b$Department of physics, Stanford University\\
382 via Pueblo Mall, Stanford CA 94305-4060, USA}

\vspace*{1cm}
\end{center}

\begin{abstract}
Using the supergravity solution of  $N_1$ D3-branes probing 
$A_{N_2-1}$ singularities we study the pp-wave limit of 
$AdS_5\times S^5/Z_{N_2}$. We show that there are two different pp-wave 
limits. One is the orbifold of the pp-wave limit of $AdS_5\times 
S^5$. In this case there is no symmetry enhancement.  
The other case is the same as the pp-wave  limit of $AdS_5\times S^5$ and 
therefore we again see the maximal supersymmetry.   
We will also identify operators in the
four dimensional ${\cal N}=2$ $SU(N_1)^{N_2}$ gauge theory which 
correspond to 
stringy excitations in the orbifolded pp-wave geometry.
The existence of the maximal pp-wave geometry indicates that there is a 
subsector of the corresponding ${\cal N}=2$ gauge theories which has 
enhanced ${\cal N}=4$ supersymmetry. We also study the pp-wave limits 
of $AdS_{7,4}\times S^{4,7}/Z_{N}$.

\end{abstract}

\end{titlepage}

\section{Introduction}

Although the maximal supersymmetric backgrounds have been known for a 
while, until the strong motivation by  AdS/CFT duality (see 
\cite{MAGOO} for a review), string theory was only formulated on the 
simpler case of ten dimensional flat background. The AdS/CFT 
correspondence, in its simplest form, states that string theory (or 
in certain limits, supergravity) on the $AdS_5\times S^5$ 
background is dual to the large $N$ ${\cal N}=4$ four dimensional SYM 
\cite{{M1},{GKP},{WIT1}}. According to this conjecture
the spectrum of string theory on $AdS_5\times S^5$ corresponds
to spectrum of single trace operators of the ${\cal N}=4$ gauge
theory. However, this correspondence has only been studied for the
supergravity modes on the $AdS_5\times S^5$ which are in one-to-one
correspondence with the chiral operators of the ${\cal N}=4$ gauge theory.

Another supergravity background which admits the maximal supersymmetry is 
the pp-wave \cite{{KG},{PPwave},{PPwaves}}. Soon after that the 
string theory on 
this background was formulated \cite{MAT1} and shown to be exactly 
solvable \cite{MAT2}. However, the observation that the supersymmetric 
pp-waves background can be understood as a certain limit of $AdS_5\times 
S^5$ geometry led the authors of \cite{BMN} 
to the conjecture that string theory on the maximally supersymmetric 
ten dimensional pp-waves has a description in terms of a certain 
subsector of the large $N$ four dimensional ${\cal N}=4$ $SU(N)$ 
supersymmetric gauge theory at weak coupling. More
precisely this subsector is parametrized by states with conformal
weight $\Delta$ carrying $J$ units of  charge under the $U(1)_{R}$ 
subgroup of the $SU(4)_{R}$ R-symmetry of the gauge theory, such that 
both $\Delta$ and $J$ are parametrically large (both scale as 
$\sqrt{N}$ in the large $N$ limit) while their difference
$\Delta-J$ is finite. 
Then, as further evidence the perturbative 
string spectrum was worked out from gauge theory side.

This procedure has also been generalized for the pp-wave limit of 
$AdS_5\times M^5$, where $M^5$ is smooth and preserves some 
supersymmetry, leading to the conjecture that for such background 
the pp-wave limit is universal and maximally supersymmetric
\cite{{GO},{IKM},{ZS}}. 
In particular focusing on the 
$AdS_5\times T^{1,1}$, it was shown that this pp-wave is 
identical to the one which appears in the 
$AdS_5\times S^5$ case. Therefore there should be a subsector of
the corresponding ${\cal N}=1$ gauge theory describing  
the string theory on the pp-wave, which has enhanced ${\cal N}=4$
supersymmetry. 

In this letter we consider non-smooth backgrounds and study their pp-wave 
limits. We consider the $AdS_5\times {\cal M}^5$ cases, where ${\cal M}^5$ 
is an orbifold of $S^5$, namely $S^5/Z_N$. The special case of 
$S^5/Z_2$ has been briefly  considered in \cite{IKM}. 
We start with  supergravity solution
of $N_1$ D3-branes in the $A_{N_2-1}$ singularities, the near horizon 
limit of which corresponds to geometry of the form
$AdS_5\times S^5/Z_{N_2}$ \cite{KS}.
We show that the orbifolded case admits two different pp-wave limits.
One is with half supersymmetries and is in fact the orbifolded 
version of maximally supersymmetric pp-wave. The other one is the 
maximal supersymmetric case. 
Therefore, the universality of the
pp-waves limit for the cases which $M^5$ is a smooth manifold \cite{IKM}
is not maintaining in the non-smooth cases. 

The article is organized as follows. In section 2, we
review how to construct the supergravity solution of  
$N_1$ D3-branes in the $A_{N_2-1}$ singularities. 
This solution can be obtained by supergravity solution of 
a M5-brane configuration in eleven dimensional supergravity
with worldvolume of the form $R^4\times \Sigma$ \cite{WIT2},
where $\Sigma$ is the Seiberg-Witten holomorphic curve.
In section 3, we shall perform the pp-wave limits of 
$AdS_5\times S^5/Z_{N_2}$ and show the existence of two pp-wave limits.
In section 4, we will consider
subsector of the ${\cal N}=2$ gauge theory corresponding
to the string on the orbifolded pp-wave. 
In section 5, we study the eleven dimensional $AdS_4\times S^7/Z_N$
and $AdS_7\times S^4/Z_N$ orbifolds and their pp-wave limits. We show that 
for the  $AdS_4\times S^7/Z_N$ case there are two pp-wave limits, one 
is the
orbifold version of eleven dimensional pp-waves of \cite {{PPwave},{BMN}} 
and the 
other is the maximal supersymmetric case. However, for the $AdS_7\times 
S^4/Z_N$ the only possible pp-wave limit is the maximally supersymmetric 
case.

\section{Gravity solution of $AdS_5\times S^5/Z_{N_2}$}

In this section we present the supergravity solution of
$AdS_5\times S^5/Z_{N_2}$ starting from intersecting M5-branes 
solution. This supergravity solution has been studied in
\cite{{FS},{AO}}.
We denote the eleven dimensional space-time coordinates by $(x_{||},
\vec{x},\vec{y},\vec{z})$,
where $x_{||}$ parameterize the $(0,1,2,3)$ coordinates,
$\vec{x}=(x_1,x_2)$ the $(4,5)$ coordinates, 
$\vec{y} = (y_1,y_2)$ the $(6,7)$ coordinates and 
$\vec{z}=(z_1,z_2,z_3)$ the $(8,9,10)$ coordinates.

Let us start with two sets of fivebranes in M theory: $N_1$ 
coinciding M5 branes and $N_2$ 
coincident M5' branes.
Their worldvolume coordinates are  $M5: (x_{||},\vec{y})$
and $M5': (x_{||},\vec{x})$. 
Such a configuration preserves eight supercharges.
The eleven-dimensional supergravity background is given by \cite{PT,T,GKT}
\be
ds^2_{11} = (H_1H_2)^{2/3}[(H_1H_2)^{-1}dx^2_{||}
+H^{-1}_2 d\vec{x}^2+H^{-1}_1 d\vec{y}^2+ d\vec{z}^2] \ ,
\label{m5m5}
\ee
with the 4-form field strength ${\cal F}$
\be
{\cal F} = 3\left(*d(H_1^{-1})\wedge dy^1 \wedge dy^2 + *d(H_2^{-1})\wedge 
dx^1 
\wedge dx^2\right) \ ,
\ee
where $*$ defines the dual form in the three dimensional space $(z_1,z_2,z_3)$.
Consider the semi-localized case when the M5 branes are completely 
localized while the
M5' branes are only localized along the overall transverse directions.
When the branes are at the origin of $(\vec{x},\vec{z})$ 
space the harmonic functions in the near core limit of the M5' branes take 
the form
\cite{Youm,L} 
\be
H_1=1+\frac{4\pi l_p^4 N_1 N_2}{(|\vec{x}|^2+
2l_p N_2|\vec{z}|)^2}, \ \ \  
H_2=\frac{l_p N_2}{2|\vec{z}|} \ .
\label{m5m5harm}
\ee

It is useful to make a change of coordinates $l_p z= (r^2
\sin^2\alpha)/2N_2, x=r \cos\alpha, 0 \leq \alpha \leq \pi/2$. The
decoupling limit of the theory is defined by 
$l_p\rightarrow 0$ while keeping $U=r/l_p^2$ and ${\vec{\hat{y}}}=
\vec{y}/l_p$ fixed. 
In this limit the metric (\ref{m5m5}) is of the form
of a {\it warped product} of $AdS_5$ and a six dimensional manifold ${\cal
M}_6$ \cite{AO} 
\be ds^2_{11} = l_p^2 (4 \pi N_1)^{-1/3}(\sin^{2/3}\alpha)
\left ( \frac{U^2}{N_2} dx^2_{||}+ \frac{4\pi N_1}{U^2} dU^2+ d{\cal
M}_6^2 \right ) \ , \label{metric11} 
\ee 
with 
\bea 
d{\cal M}_6^2 &=& 4\pi N_1
(d\alpha^2+\cos^2\alpha d\theta^2 +\frac{\sin^2\alpha}{4}d\Omega_2^2 )
+\frac{N_2}{\sin^2 \alpha}(d{\hat y}^2+{\hat y}^2d\psi^2) \ \cr 
d\Omega_2^2&=&d\gamma^2+\sin^2\gamma d\delta^2. 
\label{M}
\eea 
The four form field in this limit is given by 
\bea 
{\cal F}&=&-2\pi
N_1l_p^3 \sin^3{\alpha}\ \cos{\alpha}\ \sin\gamma\ d\alpha\wedge
d\theta\wedge d\gamma\wedge d\delta \cr & & - {1\over 2}N_2 l_p^3\
\sin\gamma\ d{\hat y}_1\wedge d{\hat y}_2\wedge d\gamma\wedge d\delta.  
\eea 
The curvature of this metric behaves like ${\cal R}\sim {1\over l_p^2
N_1^{2/3} \sin^{8/3}\alpha}$. This brane configuration can be understood
as uplifted elliptic brane system in type IIA. It consists of $N_2$
NS5-branes with the worldvolume directions (0,1,2,3,4,5) periodically
arranged in the 6-direction and $N_1$ D4-branes with worldvolume
coordinates (0,1,2,3,6)  stretched between them. The four dimensional
theory at low energies on the D4-branes worldvolume is a supersymmetric
gauge theory with 8 supercharges. In this brane set-up the R-symmetry
group $SU(2)_R \times U(1)_R$ is realized as the rotation group
$SU(2)_{8910} \times U(1)_{45}$.

The near-horizon limit of the ten dimensional metric describing the elliptic 
type IIA brane configuration is \cite{{AO},{FS}}
\be
ds^2_{10}= l_s^2 \left (\frac{U^2}{R^2}dx^2_{||}+
R^2\frac{dU^2}{U^2}+ d{\cal M}_5^2 \right ) \ ,
\ee
where
\be 
d{\cal M}_5^2 = 
R^2\bigg(d\alpha^2+\cos^2\alpha d\theta^2
+{\sin^2\alpha\over 4}d\Omega_2^2\bigg)
+{N_2^2\over {R^2 \sin^2\alpha}}d{\hat y}^2 \ ,
\label{metric10}
\ee
and $R^2=(4\pi g_sN_1N_2)^{1/2}$. In this limit the dilaton and non-zero
three and four form field strengths of the corresponding supergravity 
solution are given by
\bea
e^{\phi}&=&\frac{g_s N_2}{R\sin\alpha}\;,\;\;\;\;H_{y\gamma\delta}=
-\frac{N_2l_s^2}{2}\sin\gamma\;,\cr&&\cr
F_{\alpha\theta\gamma\delta}&=&-2\pi g_sN_1l_s^3
\sin^3\alpha\;\cos\alpha\;\sin\gamma\;.
\label{NN}
\eea
The NS-NS B field 
corresponding to the NS5-branes charge can be read from (\ref{NN}) as
\be
B_{y\delta}=\frac{N_2l_s^2}{2}(\cos\gamma-1)\;.
\ee

It is useful to make a T-duality and study these theories from Type IIB 
string theory point of view. Doing so, we obtain \cite{FS}
\be
ds^2_{10}= l_s^2 \left (\frac{U^2}{R^2}dx^2_{||}+
R^2\frac{dU^2}{U^2}+ R^2 d{\cal M}_5^2 \right ) \ ,
\label{MET}
\ee
with
\be\label{calM}
d{\cal M}^2=d\alpha^2+\cos^2\alpha d\theta^2+{\sin^2\alpha \over 4}
d\Omega^2_2
+{\sin^2\alpha\over N_2^2}[
d\chi+{N_2\over 2}(\cos\gamma-1)d\delta]^2
\ee
where $\chi={\hat y}/l_s$. 
The IIB self-dual five form is found to be\footnote{Here we are using the 
conventions of \cite{FS}.}
\be
F_{\chi\alpha\theta\gamma\delta}= 
(*F)_{\chi\alpha\theta\gamma\delta}= {3\over 20} {R^4\over 
N_2}\alpha'^2\cos\alpha\sin^3\alpha\sin\gamma \ .
\ee
The curvature of the above metric behaves like ${\cal R}\sim{1\over R^2}$.
Note that (\ref{calM}) is the metric of 
$S^5/Z_{N_2}$ orbifold  with $Z_{N_2}\subset SU(2)\subset SU(4)$. 
Therefore the solution
is nothing but D3-branes solution in the $Z_{N_2}$ orbifold.
This theory has been considered in \cite{KS} by orbifolding
of $AdS_5\times S^5$ in the context AdS/CFT correspondence. 
The theory is four dimensional ${\cal N}=2$ SYM theory with gauge group
$SU(N_1)^{N_2}$ with $N_2$ bi-fundamental matter
hypermultiplets {\it i.e.} $(Q_i, {\tilde Q}_i)$ in 
${\cal N}=1$ notation. The $Z_{N_2}$ acts as a permutation of the gauge 
factors.
The dimensionless gauge coupling of each $SU(N_1)$ part of the 
gauge group is $g_{YM}^2 \sim g_s N_2$.

\section{PP-waves as limits of $AdS_5\times S^5/Z_{N_2}$}

Following \cite{BMN}, we obtain  the pp-wave geometries arising as
limits of $AdS_5\times S^5/Z_{N_2}$ and then study string theory on 
pp-wave background through the corresponding gauge theory. 
The strategy is to 
consider the trajectory of a particle which is moving very fast along the 
$S^5/Z_{N_2}$ and focusing on the geometry near  this trajectory.
We note that the metric of $AdS_5\times S^5/Z_{N_2}$ background 
(\ref{MET}) can be recast to
\bea\label{S/ZN}
l_s^{-2}ds^2&=&R^2\bigg{[}-dt^2\cosh^2\rho+d\rho^2+\sinh^2\rho d\Omega_3^2\cr
&&\cr
&+&d\alpha^2+\cos^2\alpha d\theta^2+{\sin^2\alpha \over 4}d\Omega^2_2
+{\sin^2\alpha\over N_2^2}[
d\chi+{N_2\over 2}(\cos\gamma-1)d\delta]^2\bigg{]}
\eea 
Depending on the fact that particle trajectories we consider are near the 
singular point or far from that one can distinguish two different
interesting pp-wave limits:

\vskip .3cm  

{\it 1) Trajectories near $\alpha=0$}
\vskip .3cm  

Let us consider a particle moving along the $\theta$ 
direction and sitting at
$\rho=0$ and $\alpha=0$. We recall that $\theta$ parametrizes the 
direction which is invariant under the $Z_{N_2}$ action.
The geometry close to this
trajectory can be obtained by 
the following rescaling
\be
t=x^{+}+R^{-2}x^-\;,\;\;\;\;\; \theta=x^+ -R^{-2}x^{-}\;,\;\;\;\;\;
\rho={r\over R}\;,\;\;\;\;\;\alpha={v\over R}\;,\;\;\;\;\;R\rightarrow 
\infty\;,
\label{LIMIT}
\ee
and keeping $x^+,x^-, r$ and  $v$ fixed.  
In this limit the above metric reads as 
\bea
l_s^{-2}ds^2&=&-4dx^{+}dx^{-}-(r^2+v^2)dx^{+2}+dr^2+r^2d\Omega_3^2\cr
&&\cr 
&+&dv^2+{1\over 4}v^2\left[d\Omega^2_2+{4\over N_2^2}[d\chi+{N_2\over 
2}(\cos\gamma-1)d\delta]^2\right]\;.
\label{PPOR}
\eea
We also have a self-dual RR 5-form flux, 
$(*F)_{+v\gamma\delta\chi}=F_{+v\gamma\delta\chi}\sim {1\over 
N_2}d(Vol_4)$, where $d(Vol_4)$ is volume form for $v\gamma\delta\chi$ 
directions. Besides the explicit form of metric (\ref{PPOR}), the fact 
that the five form flux is proportional to ${1\over 
N_2}$ justifies that our geometry contains the $Z_{N_2}$ 
orbifold. 
Since the curvature of the above metric is proportional to 
${1\over R^2}$,  in the $R\to\infty$ limit  curvature always remains 
small.
This pp-wave limit of $AdS_5\times S^5/Z_{N_2}$ is 
$Z_{N_2}$ orbifold of the pp-wave limit of $AdS_5\times S^5$
\cite{BMN} and hence this background preserves only half of 
supersymmetry (16 supercharges).

\vskip.3cm
{\it 2) Trajectories near $\alpha={\pi\over 2}$}
\vskip.3cm
Besides the above limit there is another pp-wave metric considering the 
particles moving  very fast along the $\chi$ direction, sitting at 
$\rho=0$, $\gamma=0$ and $\alpha={\pi\over 2}$. Note that $\chi$ is 
the direction involved in  the orbifolding. Let us consider the 
scaling
\bea\label{limit2}
t=x^+ +R^{-2}x^-, \ \ \ \ \ \ \   {1\over N_2}\chi=x^+ -R^{-2}x^-, \cr  
\rho={r\over R}\ ,\ \ \gamma={2x\over R}\ ,\ \ \ \alpha={\pi\over 
2}-{y\over R}\ ,
\ R\to\infty 
\eea
while keeping $x^+$, $x^-$, $r,\ x$ and $y$ fixed.
In this limit the metric (\ref{S/ZN}) becomes
\be
l_s^{-2}ds^2=-4dx^{+}dx^{-}-\mu^2{\vec z}^2dx^{+2}+d{\vec z}^2\ ,
\label{MAX}
\ee
and
\be
F_{+1234}=F_{+4567}={3\over 20}\mu\ .
\ee
To obtain the above metric we have redefined 
$\delta-x^+$ as the new coordinate $\delta$. 
Moreover the mass parameter $\mu$ has been introduced by
reacaling $x^{-}\rightarrow x^{-}/\mu$ and  
$x^{+}\rightarrow \mu x^{+}$.
As we see the above metric, as well 
as the self dual five form field, do not 
depend on $N_2$ and it is exactly the 
pp-wave limit of $AdS_5\times S^5$ discussed in \cite{BMN}. 
This pp-wave limit corresponds to a maximally 
supersymmetric background. The very fact that the $N_2$ dependence has 
been removed in the pp-wave limit metric can also be understood from 
the four dimensional gauge field theory side. For that it is enough to 
recall that in the scaling limit (\ref{limit2}) $x^+$ is related to $\chi$ 
by a factor of $N_2$ and hence the $U(1)_R$ charges of the 
states should now be measured in units of ${N_2}$.  

One can work out the pp-wave limit of the
eleven dimensional solutions of (\ref{metric11}), (\ref{M}). 
This we have presented 
in the Appendix.  We note that the  
maximally supersymmetric pp-wave do not have an eleven dimensional 
counter-part. 

Therefore, for the $AdS_5\times {\cal M}^5$ spaces, where ${\cal M}^5$ is 
an orbifold of 
$S^5$ which preserves some supersymmetry, depending on the scaling we use
for going to pp-wave limit, one may find maximally supersymmetry case or
the cases with (half) of supersymmetry broken.

Here we will mainly focus on the first pp-wave case. Following 
\cite{BMN} we can write the light-cone momenta in terms of
conformal weight $\Delta$ and the angular momentum, 
$J=-i\partial_{\theta}$, of the operator in the
superconformal field theory as following
\bea
2p^{-}&=&i\partial_{x^+}=i(\partial_t+\partial_{\theta})=\Delta-J
\cr &&\cr
2p^{+}&=&i\frac{\partial_{x^-}}{R^2}=i
\frac{(\partial_t-\partial_{\theta})}{R^2}=\frac{\Delta+J}{R^2}\ .
\eea
Configurations with fixed non-zero $p^{+}$ in the limit (\ref{LIMIT}) 
correspond to the states in the $AdS$ with large 
angular momentum $J\sim R^2$. In the gauge theory side such 
configurations correspond to operators with R-charge $J\sim \sqrt{N_1}$ 
and  $\Delta-J$ fixed, in the $N_1\rightarrow \infty$ limit
and keeping  the gauge theory coupling $g^2_{YM}$, fixed and small.
In next section we will study these operators and identify them
with excited string states.

On the other hand for the pp-wave with maximal supersymmetry (\ref{MAX})
the light-cone momenta in terms of
conformal weight $\Delta$ and the angular momentum 
$J=-i\partial_{\chi}$ can be written as
\bea\label{maxp}
2p^{-}&=&i\partial_{x^+}=i(\partial_t+N_2\partial_{\chi})=\Delta-N_2J
\cr &&\cr
2p^{+}&=&i\frac{\partial_{x^-}}{R^2}=i
\frac{(\partial_t-N_2\partial_{\chi})}{R^2}=\frac{\Delta+N_2J}{R^2}\ .
\eea
Therefore in this case we will be looking for spectrum of states with 
$\Delta-N_2J$ finite in the limit $N_1\rightarrow \infty$. In fact from
pp-wave limit (\ref{MAX}) we learn that there should  also be a subsector
of ${\cal N}=2$ which exhibits the supersymmetry enhancement, similar
to the conifold case \cite{{GO},{IKM}}. This subsector is parametrized
by $\Delta-N_2J$ eigenvalue. We would like to comment that, since the above  
maximal SUSY PP-wave limit do not depend on $N_2$, one can study the large 
$N_2$ limit. In particular equations (\ref{maxp}) are very suggestive that
in the large $N_2$ one has the chance to keep $J$ finite. However, here we
will only consider the half SUSY case and in the next section we construct 
the gauge theory operators corresponding to the strings on PP-wave 
orbifolds. Giving the description of strings in PP-waves in terms of 
${\cal N}=2$ gauge theories is very interesting and important question which 
we will come back to, in the future works \cite{Decons}.

\section{Strings in PP-wave orbifolds from ${\cal N}=2$ SYM } 

The twisted and untwisted states of AdS orbifolds have been studied
in \cite{{OT},{G}}. In fact the twisted states of $AdS_5\times S^5/
N_{2}$ orbifold can be identified with the  chiral primary 
operators of ${\cal N}=2$ $SU(N_1)^{N_2}$ conformal SYM theory which 
are not invariant under exchange of the gauge group factors, while
the untwisted states can be identified with those which
are invariant. In our notation the untwisted operators can be considered 
as single trace operators which are symmetric under exchange of gauge
group, {\it i.e.} $\sum_{i=1}^{N_2}{\rm Tr({\varphi}_i^k)}$, 
where ${\varphi}_i$'s 
are adjoint complex scalars in the vectormultiplet
of each gauge factor and carry a unit of R-charge  under $U(1)_{R}$ 
factor of 
R-symmetry group. This operator has conformal weight $k$. 

On the other hand the twisted operators which are not invariant under
exchange of the gauge factors may be written in the following form
\cite{G}
\be
{\cal O}(i)-{\cal O}(i+1)\;,
\label{TWIST}
\ee  
where ${\cal O}$ can be thought as a basis which is constructed from a
certain combination of chiral fields. For example we can consider those
operators which are given in terms of scalars in the vectormultiplet,
{\it i.e.} ${\cal O}(i)={\rm Tr({\varphi}_i^k)}$. 

We want to study those states which carry large R-charge
$J\sim\sqrt{N_1}$
under the generator of $U(1)_{R}$ subgroup of the 
R-symmetry which acts on the adjoint scalars in the vectormultiplet
while has no effect on the scalars of the bi-fundamental matters; and
we are looking for the spectrum of
states with $\Delta-J$ fixed. 
In the untwisted sector the operator with lowest value of 
$\Delta-J$, which is zero, is $\sum_{i=1}^{N_2}{\rm Tr({\varphi}_i^J)}$.
This operator is chiral primary and therefore its dimension
is protected by supersymmetry and, is associated to 
the untwisted vacuum state of the corresponding string theory in the 
light-cone gauge, much 
similar to the ${\cal N}=4$ case \cite{BMN}. Operators in the twisted 
sector with 
lowest value of $\Delta-J$ can be read from (\ref{TWIST}) by
setting ${\cal O}(i)={\rm Tr({\varphi}_i^J)}$.

Now we can proceed with constructing other operators.\footnote{Here
we shall only consider the bosonic operators. Their
superpartners can be easily obtained using  
the ${\cal N}=2$ supersymmetry.} These
operators can be parameterized by their $\Delta-J$ 
eigenvalue.  For example for $\Delta-J=0$ the
corresponding operators are just what we have presented
above. According to the prescription given in \cite{BMN} the 
other operators can be obtained by inserting new 
operators in the trace. In other words replacing some of
$\varphi_i$ in the operator ${\rm Tr({\varphi}_i^J)}$ by
other fields. 

The bosonic operators which can be inserted 
in the trace are as following. 
We have four scalars in the directions that are not
rotated by $J$. They are, in fact, the scalars in the
hypermultiplet  which correspond to the directions
in which the $Z_{N_2}$ orbifold is defined. On the
other hand there are  four other fields which are
made out of ${\varphi}_i$.  These fields are derivative
of ${\varphi}_i,\; D_{a}{\varphi}_i+[A_a,{\varphi}_i]$, where
$a=1,2,3,4$ are the directions in $R^4$ parameterized by
${\vec r}$ in (\ref{PPOR}). Since these directions are
not affected by the $Z_{N_2}$ orbifold, they
should lead to the same excitation as those in
$AdS_5\times S^5$. These are operators with $\Delta-J=1$.

The next bosonic operators we shall consider are those with
$\Delta-J=2$. A basis for these operators can be obtained
 by inserting the scalars in the hypermultiplet {\it
i.e.} $(Q^{\mu}_i,{\tilde Q}^{\mu}_i)$ in 
trace ${\rm Tr({\varphi}_i^J)}$ for $\mu=1,2$. These
scalars which correspond to the
directions where the orbifold is defined, are in the
bi-fundemantal representation of the gauge group.
This means that under a gauge transformation they 
transform as $Q_i^{\mu}\rightarrow U_i Q_i^{\mu}U^{-1}_{i+1}$ and
${\tilde Q}_i^{\mu}\rightarrow U_{i+1} {\tilde Q}_i^{\mu}U^{-1}_{i}$.
Being bi-fundemantal scalars, we need to insert at least two of 
them in the trace to make a gauge invariant opeartor. 
Since these scalars 
are affected  by the $Z_{N_2}$ orbifold, we can construct 
operators both in the untwisted (and twisted) sectors by
using a combination of these basis which are invariant 
(and non-invariant) under exchange of the gauge group factor. 
The basis for these gauge invariant operators are given by
\bea
{\cal O}^n_1(j)&=&\sum_{l=1}^J{\rm Tr}\left(\varphi_j^l\,Q^{\mu}_j
\,\varphi_{j+1}^{J-l}\,{\bar Q}^{\nu}_j\right)\;e^{2\pi i nl\over N_2J},
\cr &&\cr
{\cal O}^n_2(j)&=&\sum_{l=1}^J{\rm Tr}\left(\varphi_j^l\,Q^{\mu}_j
\,\varphi_{j+1}^{J-l}\,{\tilde Q}^{\nu}_j\right)\;e^{2\pi i nl\over N_2J},
\cr &&\cr
{\cal O}^n_3(j)&=&\sum_{l=1}^J{\rm Tr}\left(\varphi_j^l\,
{\bar {\tilde Q}}^{\mu}_j\,
\varphi_{j+1}^{J-l}\,{\tilde Q}^{\nu}_j\right)\;e^{2\pi i nl\over N_2J},
\cr &&\cr
{\cal O}^n_4(j)&=&\sum_{l=1}^J{\rm Tr}\left(\varphi_j^l\,
{\bar {\tilde Q}}^{\mu}_j\,
\varphi_{j+1}^{J-l}\,{\bar Q}^{\nu}_j\right)\;e^{2\pi i nl\over N_2J}.
\label{DELJ2}
\eea
Analogous to \cite{BMN} these operators with $n$ being an integer multiple 
of $N_2$ correspond 
to the excited string states obtained by applying the creation 
operators $a_{-n/N_2}^{\mu}(j) a_{-n/N_2}^{\nu}(j+1)$ on the 
untwisted or twisted 
light-cone vacuum. For $n$ not an integer multiple of $N_2$ we have just 
states in twisted sector.

In general one can construct gauge invariant operators with 
$\Delta-J=2(1+k)$ for $0\leq k\leq N_2-1$. These operators are obtained
by inserting 
$k+1$ scalars in the hypermultiplet into the trace in such a way that the
whole combination is gauge invariant. For example
one can consider the following operator with $\Delta-J=2(1+k)$ 
\bea
{\cal O}_1^n(j)&=&\sum_{l=1}^J{\rm Tr}\left(\varphi_j^l\;Q^{\mu_0}_j
\cdots Q^{\mu_k}_{j+k}
\;\varphi_{j+k+1}^{J-l}\;{\bar Q}^{\nu_k}_{j+k}\cdots {\bar Q}^{\nu_0}_j
\right)\;e^{2\pi i nl\over N_2J},\cr 
&&\cr
{\cal O}_2^n(j)&=&\sum_{l=1}^J{\rm Tr}\left(\varphi_j^l\;Q^{\mu_0}_j
\cdots Q^{\mu_k}_{j+k}
\;\varphi_{j+k+1}^{J-l}\;{\tilde Q}^{\nu_k}_{j+k}\cdots 
{\tilde Q}^{\nu_0}_j
\right)\;e^{2\pi i nl\over N_2J},\cr 
&&\cr
{\cal O}_3^n(j)&=&\sum_{l=1}^J{\rm Tr}\left(\varphi_j^l\;
{\bar {\tilde Q}}^{\mu_0}_j\cdots {\bar {\tilde Q}}^{\mu_k}_{j+k}\;
\varphi_{j+k+1}^{J-l}\;{\tilde Q}^{\nu_{k}}_{j+k}\cdots
{\tilde Q}^{\nu_0}_{j}\right)\;
e^{2\pi i nl\over N_2J}\cr 
&&\cr
{\cal O}_4^n(j)&=&\sum_{l=1}^J{\rm Tr}\left(\varphi_j^l\;
{\bar {\tilde Q}}^{\mu_0}_j\cdots {\bar {\tilde Q}}^{\mu_k}_{j+k}\;
\varphi_{j+k+1}^{J-l}\;
{\bar Q}^{\nu_{k}}_{j+k}\cdots
{\bar Q}^{\nu_0}_{j}\right)\;
e^{2\pi i nl\over N_2J}
\eea
The corresponding string state can be obtained by applying  
creation operators like $a_{-n/N_2}^{\mu_0}(j)\cdots 
a_{-n/N_2}^{\mu_k}(j+k) a_{-n/N_2}^{\nu_0}(j)
\cdots a_{-n/N_2}^{\nu_{k}}(j+k)$  on the light-cone vacuum.

\section{PP-wave limits of orbifolds of $AdS_{4,7}\times S^{7,4}$}

In section 3, studying the pp-wave limit of the $AdS_5\times S^5/Z_N$ 
orbifold we showed that depending on how we take the pp-wave limit, we 
can find maximal and half supersymmetric pp-wave backgrounds.
In this section we study the similar limits for eleven dimensional 
orbifold cases. However, first we need to work out the metric
for such orbifold solutions. These solutions have been studied in
\cite{Pelc}.

To obtain the supergravity 
solution corresponding to $AdS_{4,7}\times S^{7,4}/Z_N$ 
supersymmetric orbifolds, 
one may start with M2 or M5 -branes probing the orbifold. 
\footnote{ The corresponding field theory in these cases are ${\cal 
N}=(0,1)$ six 
dimensional SCFT  for $AdS_7\times S^4$ orbifold and 
three dimensional ${\cal N} = 2$ SCFT in the Large $N$ limit for the 
$AdS_4\times S^7$ orbifold \cite{Radu}.}
For that 
it is enough to replace the metric for the transverse direction by the 
proper orbifold solution. Since the orbifolds are Einstein manifolds, 
it is  guaranteed that solutions found in this way are supergravity 
solutions.
As we have  shown in section 3, the metric of a supersymmetric $S^3$  
orbifold, {\it i.e.} $S^3/Z_{N}$ is given by
\be
d{\tilde \Omega}_3^2={1\over 4}d\Omega_2^2+{1\over N^2}\left[d\chi+{N\over
2}(\cos\gamma-1)d\delta\right]^2\;,
\label{ORB}
\ee
where $d\Omega_2^2=d\gamma^2+\sin^2\gamma d\delta^2$. Here $0\leq \gamma
\leq \pi$ and  $\delta$ and $\chi$ are periodic with period $2\pi$. In 
fact
the metric (\ref{ORB}) parameterizes the unit 3-sphere where $\chi$ ranges
from 0 to $2\pi N$. The $Z_{N}$ action which leads to an orbifold is
given by $\chi \equiv \chi+2\pi$. 
With this information, we are ready to write down the eleven dimensional 
$AdS$ orbifold solutions. 

\subsection{The pp-wave limit of $AdS_4\times S^7/Z_N$}
 
Let us consider the $AdS_4\times S^7$ metric\footnote{In our notations $R$ 
is the radius of the $AdS$ piece which is half of the radius of the 
$S^7$ for $AdS_4$ case and is twice bigger than the radius of $S^4$ 
for the $AdS_7$ case.} 
\bea\label{AdS4}
l_p^{-2}ds^2=R^2\left(-\cosh^2\rho dt^2+d\rho^2+\sinh^2\rho 
d\Omega^2_2+ 4 d\Omega^2_7\right).
\eea
Now we split the transverse direction to two parts
\be
d\Omega_7^2=d\alpha^2+\cos^2\alpha \;d{\Omega'}_3^2+\sin^2\alpha\; 
d\Omega_3^2\;.
\ee
Then replacing $d\Omega_3^2$ with $d{\tilde \Omega}_3^2$,
we obtain
\be\label{S7/ZN}
d\Omega_7^2=d\alpha^2+\cos^2\alpha \;d{\Omega'}_3^2+\sin^2\alpha\;
d{\tilde \Omega}_3^2\;,
\ee
which is the metric for $S^7/Z_N$. Finally the 
$AdS_{4}\times S^7/Z_{N}$ metric is obtained by replacing the 
the $S^7$ part of (\ref{AdS4}) with (\ref{S7/ZN}).
The above can also be understood as the near horizon geometry 
of the supergravity solution of $N_1N$ M2-branes in the orbifold 
background, where $R=(\pi^2 N_1N/2)^{1/6}$. The above solution can also be 
understood as T-dual of the near horizon limit
of semi-localised intersection of the D3/D5/NS5 system \cite{Pope}.

To obtain the pp-wave limits we note that the orbifold action on $S^7$ 
leaves a $S^3$ part invariant. Now, we can proceed in the same lines of 
section 3. Again there are two possibilities 
corresponding to two following choices:
\vskip .3cm

{\it 1)  $t=x^++R^{-2}x^-\;\;\;\;\;\theta=x^+-R^{-2}x^-,
\;\;\;\;\beta={w\over R}$}
\vskip .3cm
\noindent
where $\theta$ and $\beta$ are 
directions along the 
invariant $S^3$.\footnote{Here we define
$d{\Omega'}_3^2=\cos^2\beta\;d\theta^2+d\beta^2+\sin^2\beta\; d\eta^2$}
The other coordinats are scaled as (\ref{LIMIT}).
This case, will correspond to the orbifold of the maximally supersymmetric 
pp-wave discussed in \cite{{PPwave},{BMN}} and hence half of 
supersymmetric is broken.
\vskip .3cm

{\it 2)  $t=x^++R^{-2}x^-\;\;\;\;\;{\chi\over N_2}=x^+-R^{-2}x^-$},
\vskip .3cm
\noindent 
and the other limits as defined in (\ref{limit2}).
In this case we will end up with the maximal supersymmetric eleven 
dimensional pp-wave background.

\subsection{The pp-wave limit of $AdS_7\times S^4/Z_N$}

One can repeat the above discussions for the  $AdS_7\times S^4/Z_N$
case (or equivalently the near horizon geometry of M5-brane probing the 
orbifold). The metric for $S^4/Z_N$ is
\be
d\Omega_4^2=d\alpha^2+\sin^2\alpha\; d{\tilde \Omega_3}^2\;.
\ee
Hence, the supergravity metric for the  $AdS_7\times S^4/Z_N$ will be 
given by 
\bea\label{AdS7}
l_p^{-2}ds^2&=&R^2\biggl\{-\cosh^2\rho dt^2+d\rho^2+\sinh^2\rho 
d\Omega^2_5 \cr
& &+ {1\over 4}d\alpha^2+{1\over 16}{\sin^2\alpha} 
\left(d\Omega_2^2+{4\over 
N^2}[d\chi+{N\over
2}(\cos\gamma-1)d\delta]^2\right)\biggr\}\;.
\eea 
where $R=2 (\pi N_1N)^{1/3}$.
In this case, unlike the previous cases, there is no invariant direction 
in the $S^4$ part of the metric. Therefore, the only pp-wave limit will 
correspond to the maximally supersymmetric case \cite{PPwave} which is 
obtained  
through
\be 
t=x^++R^{-2}x^-\;\;\;\;\;\;{\chi\over 2N}=x^+-R^{-2}x^-, \ \ \ R\to\infty,
\ee
limit (the other coordinates scaled similar to (\ref{limit2})).

\section{Discussions}

In this note we have studied the Penrose limit of orbifolded $AdS$ 
geometries. Considering $AdS_p\times S^q/Z_N$ cases we have shown that 
for $q>4$ in general there are two different pp-wave limits, our result
can be summarized as follows:
 
Consider  ${\cal M}/Z_N$ orbifold, which preserves half supersymmetries. 
In general the orbifolding acts
only on  some subgroup of the isometries of ${\cal M}$ and leaving some
subgroups  invariant. Now, if we study the geometries near the
trajectories along the invariant
directions, we will find half supersymmetries. However, if we consider 
the  
directions which are not invariant under the orbifold direction we may   
find the maximal supersymmetric case.

In particular for the $AdS_7\times S^4$ case, the isometries are $SO(5)$ 
and the $Z_N$ acts of a $SU(2)$ subset, 
there is no invariant direction. Therefore, we do not find half 
supersymmetric case.

We have also discussed how to obtain the strings on the $Z_{N_2}$ orbifold 
of pp-wave limit of $AdS_5\times S^5$ from the corresponding gauge theory, 
which in this case is large $N_1$ ${\cal N}=2,\ D=4$ $SU(N_1)^{N_2}$  
theory at weak Yang-Mills (but large 't Hooft) coupling. The relevant 
operators in the gauge 
theory are those with large conformal weight $\Delta$ and large R-charge 
$J$, with $\Delta-J$ fixed. In particular we have identified $\Delta-J=0$
(the corresponding string theory vacuum) and $\Delta-J=1,2$ operators, 
which correspond to higher order stringy oscillations. 
The twisted and untwisted states of the stringy excitations 
correspond to a subsector of ${\cal N}=2$ operators
which are non-invariant and invariant under the exchange of the gauge
group factors, respectively.
The question of constructing the strings in (maximal SUSY) PP-waves in terms 
of the ${\cal N}=2$ operators is an interesting and important question and
will be studied in  future works \cite{Decons}.

It would be very interesting to study M-theory on the pp-wave 
limit of $AdS_4\times S^7/Z_N$ which preserves 16 supersymmetries. Similar 
to the maximal supersymmetric case, we expect that to have a Matrix 
description.  We postpone this to future works.

\vskip .5cm

{\bf Acknowledgments}

We would like to thank Keshav Dasgupta for discussions and Sunil Mukhi for 
helpful comment.
The research of M. M. Sh-J. is supported in part by NSF grant PHY- 9870115 
and in part by funds from the Stanford Institute for Theoretical Physics.

\vskip 1cm

{\large\bf{Appendix}}

\vskip .5cm

In the same spirit, one can also study the pp-wave limits of eleven 
dimensional solution (\ref{metric11}), (\ref{M}).
For this purpose it is more convenient to write the $AdS_5$ piece in the 
global $AdS$ coordinates:
\bea
l_p^{-2}ds^2&=&({R^4\sin\alpha\over 
N_2})^{2/3}\bigg{[}-dt^2\cosh^2\rho+d\rho^2+\sinh^2\rho d\Omega_3^2\cr
&&\cr
&+&d\alpha^2+\cos^2\alpha d\theta^2+{\sin^2\alpha \over 4}d\Omega^2_2
+{N_2^2\over R^4\sin^2\alpha}(d{\hat y}^2+{\hat y}^2d\psi^2)\bigg{]}
\nonumber
\eea 
where $R^4=4\pi N_1 N_2$.  In the limit
\be\ba {ll}
t=x^+ +R^{-2}x^-, &\;\;\;\; \theta=x^+ -R^{-2}x^-, \cr  
\rho={r\over R},&\;\;\;\; \alpha={v\over R}\ ,
\;\;\; R\to\infty \; ,
\label{limit11}
\nonumber
\ea\ee
and  keeping $x^+$, $x^-$, $r$ and $v$ fixed. In this limit the metric 
becomes
\bea
l_p^{-2}ds^2&=&({v\over 
N_2})^{2/3}\bigg{[}-4dx^+dx^- -({\vec r}^2+v^2)(dx^+)^2+d{\vec r_4}^2
\cr &&\cr &+&(dv^2+{1\over 4}v^2d\Omega_2^2)
+{N_2^2\over v^2}(d{\hat y}^2+{\hat y}^2d\psi^2)\bigg{]}.
\nonumber
\eea 
The four form field in the above limits becomes
\bea
l_p^{-3}{\cal F}_{+v\gamma\delta}&=&-{1\over N_2} v^3\ \sin\gamma \cr 
l_p^{-3}{\cal F}_{{\hat y}_1{\hat y}_2\gamma\delta}&=&-{N_2\over 
2}\sin\gamma\ . 
\nonumber
\eea
In this limit the curvature reads as $l_p^2{\cal R}\sim 
N_2^{2/3}/v^{8/3}$.
Therefore one can trust the gravity for large $v$. The singularity at 
$v=0$
can be interpreted as a signal that some degrees of freedom have been
effectively integrated and are needed in order to resolve the singularity.
The (DLCQ) of M-theory on the above background should correspond to a 
Matrix-theory. However, we do not study that Matrix theory here.

\end{document}